\documentclass[twocolumn,aps,floats,superscriptaddress,showpacs]{revtex4}
\usepackage{amsmath}

\usepackage{epsfig}
\usepackage{graphicx}
\usepackage{dcolumn}
\usepackage{bm}

\def\gapp{\lower.35em\hbox{$\stackrel{\textstyle>}{\sim}$}}
\def\lapp{\lower.35em\hbox{$\stackrel{\textstyle<}{\sim}$}}

\begin{document}
\bibliographystyle{apsrev}
%

\title{Effect of Coulomb interactions on
the optical properties  of  doped graphene}

\author{ Adolfo G. Grushin, Bel\'en Valenzuela, and Mar\'{\i}a A. H. Vozmediano}

\affiliation{Instituto de Ciencia de Materiales de Madrid,\\
CSIC, Cantoblanco, E-28049 Madrid, Spain.}

\date{\today}
\begin{abstract}
Recent optical conductivity experiments of doped graphene
in the infrared regime reveal a strong background in the energy
region between the intra and interband transitions difficult to
explain within conventional pictures. We propose a
phenomenological model taking into account the marginal Fermi
liquid nature of the quasiparticles in graphene near the
neutrality point that can explain qualitatively the observed
features. We also study the electronic Raman signal and suggest
that it will  also be anomalous.

\end{abstract}
%
%
%
%

\maketitle

{\it Introduction.}
The role of many body corrections to the physics of graphene
is at this point uncertain. While in the first transport
experiments electron-electron interactions seemed  not to play a
major role \cite{Netal04,Netal05,Zetal05}, more recent measurements
indicate the possible importance of many body corrections to the
intrinsic properties of graphene
\cite{Jetal07,Oetal07,ZSetal08,LLA09,NBetal08,MSetal08,Letal08}.
The electronic transport of single layer graphene is one of the
most intriguing aspects  that still lacks a full understanding.
Unlike common Fermi liquids, graphene shows  a minimal
conductivity at zero frequency in the neutral suspended samples
with a value of the order of the conductivity quantum, $e^{2}/h$
that grows linearly with the electron density in the low density
regime \cite{Netal05,Netal06,KNG06}.

The optical properties of graphene (see \cite{GSC07} for a very
accurate review) have been the focus of interest of recent
experiments \cite{NBetal08,MSetal08,Letal08,Ketal08}. In the visible
region they allow a very precise determination of the value of
the fine structure constant. The structure of graphene is similar to
that of a two-dimensional two band semiconductor. The dynamical
conductivity is expected to have a characteristic shape where
intraband  (near zero energy) and interband (at an energy of twice
the chemical potential, $2\mu$) transitions can be clearly
identified. Recent observations in the infrared region of the
spectrum \cite{Letal08}, reveal a substantial background of
approximately 30 percent of the saturation value in the frequency
range between intraband and interband transitions. The observed
background remains approximately constant for several values of the
chemical potential that range from $0.10$ eV up to $0.30$ eV.
Moreover, the threshold at $2\mu$ shows a very wide width
independent of gate voltage and therefore of carrier density.

A strong background was also observed in the early times of
high-$T_c$ superconductors  and prompted the marginal Fermi liquid
ideas \cite{Vetal89}. Together with the observation of the anomalous
infrared optical conductivity in \cite{Letal08}, they observe an
increase of the Fermi velocity with lowering the frequency at low
energies which is suggested to be due to the renormalization of the
Fermi velocity from the electron-electron interactions as predicted
in early works \cite{GGV94,GGV99}. The measures of optical
properties as a function of the frequency for varying chemical
potential \cite{MSetal08,Ketal08}, angle resolved spectroscopy
(ARPES) \cite{Betal07,Oetal07,ZSetal08} or scanning tunneling
microscopy (STM) \cite{LLA09} are complementary probes to explore
the influence of disorder and interactions on the properties of
graphene. In this work we propose that the mid infrared residual
conductivity found in the experiments at finite chemical potential
can be a footprint of the anomalous behavior of the electron
lifetime at the Dirac point \cite{GGV96}. We compute the optical
conductivity with a phenomenological self-energy based on the
marginal Fermi liquid behavior discussed in \cite{GGV96} and see
that it helps to understand the experiments. We predict a similar
anomalous background in the electronic contribution to the
Raman signal which has not been measured yet in graphene.
The influence of
electron-electron interactions on the conductivity of undoped
graphene was addressed in \cite{M07,SS09}.
\\
\begin{figure*}
\begin{minipage}{.49\linewidth}
\begin{center}
\includegraphics[scale=0.355]{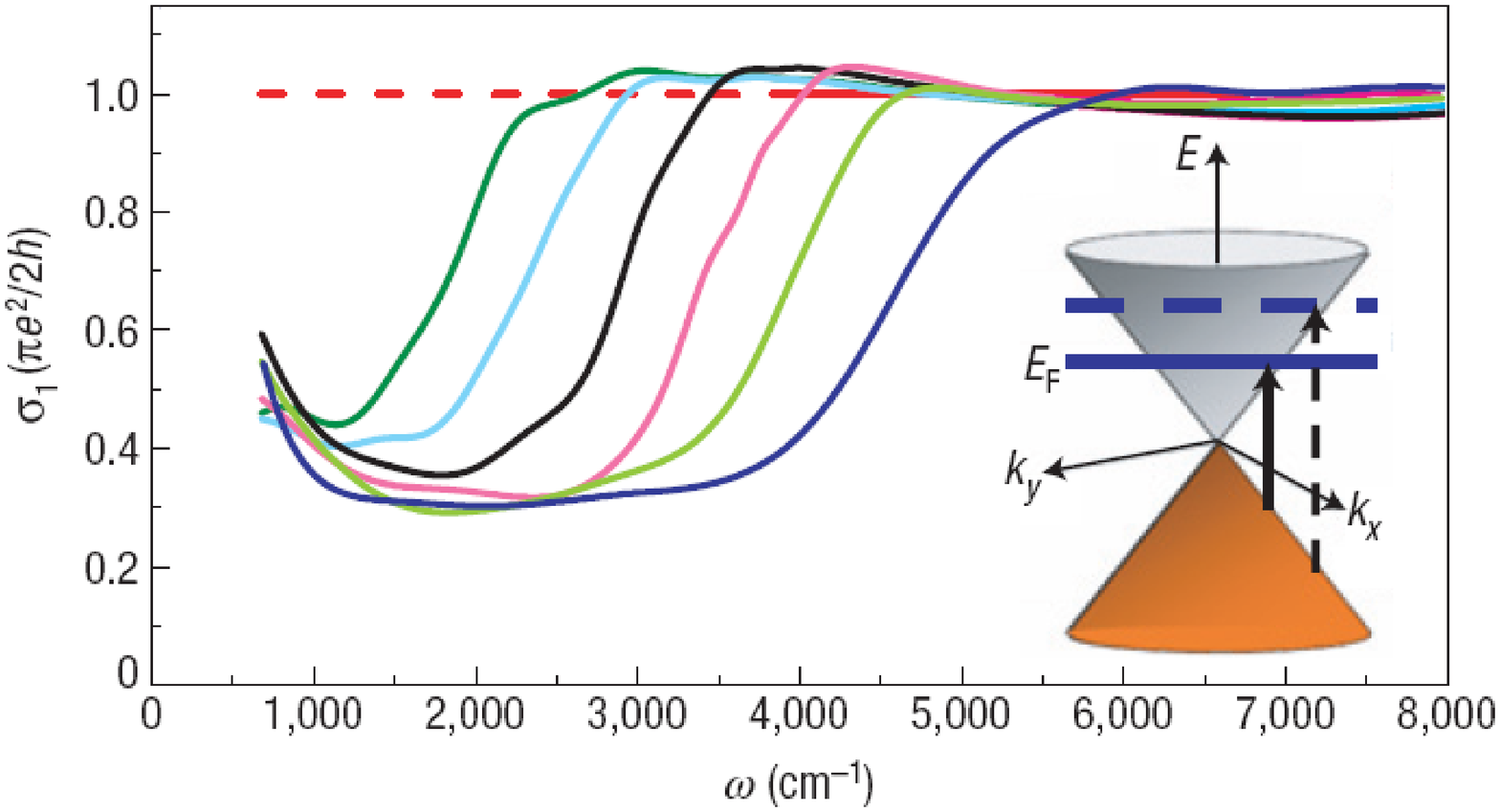}
\begin{center}
(a)
\end{center}
\end{center}
\end{minipage}
\begin{minipage}{.49\linewidth}
\begin{center}
\includegraphics[scale=0.75]{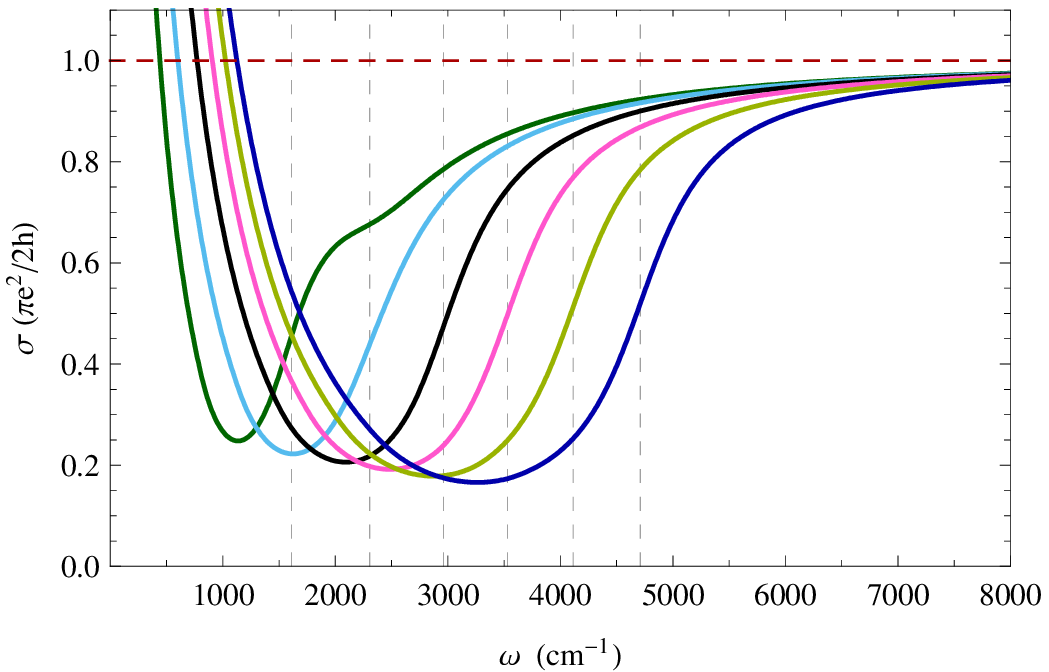}
\begin{center}
(b)
\end{center}
\end{center}
\end{minipage}\\
\caption{\label{Basov} (Color online) (a) Measured optical
conductivity of graphene for different gate voltages (from right
to left $V_{g}=$ $71$, $54$, $40$, $28$, $17$ and $10$ V). The
dashed red line is the optical conductivity at zero chemical potential.
 (b) Conductivity
computed with the self-energy (\ref{TwopartSE}) for parameters
$a=0.2$, $b=0.001$ eV, $\Lambda=0.15$ eV and for the same gate
voltages. The vertical dashed lines indicate the $2E_{F}$
threshold at each chemical potential.}
\end{figure*}
{\it Main electronic properties of graphene.}
As it is well known by now, graphene is a two dimensional array of
carbon atoms with an $sp^2$ coordination. An updated review of its
properties can be found in \cite{NGetal09}. The strong $sp^2$
sigma bonds form a honeycomb lattice and the low energy electronic
structure comes from the $\pi$ electrons perpendicular to the
plane located at the sites of the honeycomb structure. Due to the
special topology of the honeycomb lattice that has two atoms per
unit cell and lacks inversion symmetry, the Fermi surface of the
neutral system consists of two inequivalent Fermi points
\cite{W47} and the low energy excitations around them are
described in the continuum limit by the two dimensional massless
Dirac Hamiltonian \cite{S84}. The low energy  dispersion relation
of the two bands around a Fermi point is $$\epsilon({\bf k})=\pm
v_F\vert{\bf k}\vert,$$ from where it can be seen that the density
of states vanishes at the Fermi energy and  the effective mass of
the quasiparticles is zero. $v_F$ is the Fermi velocity estimated
to be $v_F\sim 10^6 m/s\sim c/300$.

The low energy nature of the interacting system was studied with
renormalization group techniques in \cite{GGV94}. Due to the
vanishing of the density of states at the Fermi energy the Coulomb
interaction is unscreened and the neutral system departs from the
usual Fermi liquid behavior. In particular the inverse lifetime of
the quasiparticles was found in \cite{GGV96} to decrease linearly
with the frequency instead of the squared Fermi liquid behavior.
\\

{\it Phenomenological model for the electron self-energy.} The
main idea of this work is that the physics of the Dirac point
influences the infrared  properties of the doped system. This
assumption has some experimental support. In the photoemission
analysis done in \cite{Betal06} it is shown that interactions
deform the spectral function of doped graphene not only at the
Fermi level but also at the position of the Dirac cone. In the
early work \cite{GGV96} it was shown that the inverse lifetime of
intrinsic graphene grows linearly with energy at low energies,
an infrared behavior typical of the marginal Fermi liquid
\cite{Vetal89}. The inverse lifetime of electrons in the presence
of the unscreened Coulomb interaction was found to be:
\begin{equation}
\label{VozSE} \mathrm{Im}\;\Sigma(\omega) = \dfrac{1}{48}\left(
\dfrac{e^{2}}{\epsilon_{0}\hbar v_{F}}\right)^{2}|\omega |.
\end{equation}
This result has been corroborated by recent experiments
\cite{LLA09} that provide an estimation of the coefficient of the
linear scattering rate of the order of 0.07 in a range of
energies up to $\sim 0.15$ eV \footnote{We thank Eva Andrei for
providing this estimation to us.}.

\begin{figure*}
\begin{minipage}{.49\linewidth}
\begin{center}
\includegraphics[scale=0.75]{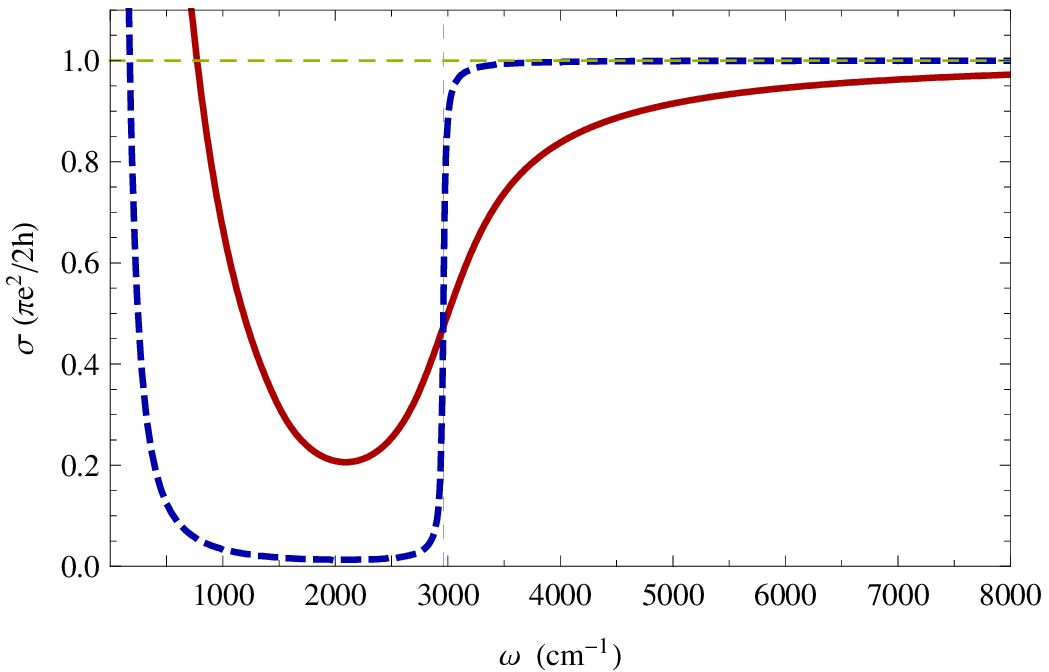}
\begin{center}
(a)
\end{center}
\end{center}
\end{minipage}
\begin{minipage}{.49\linewidth}
\begin{center}
\includegraphics[scale=0.75]{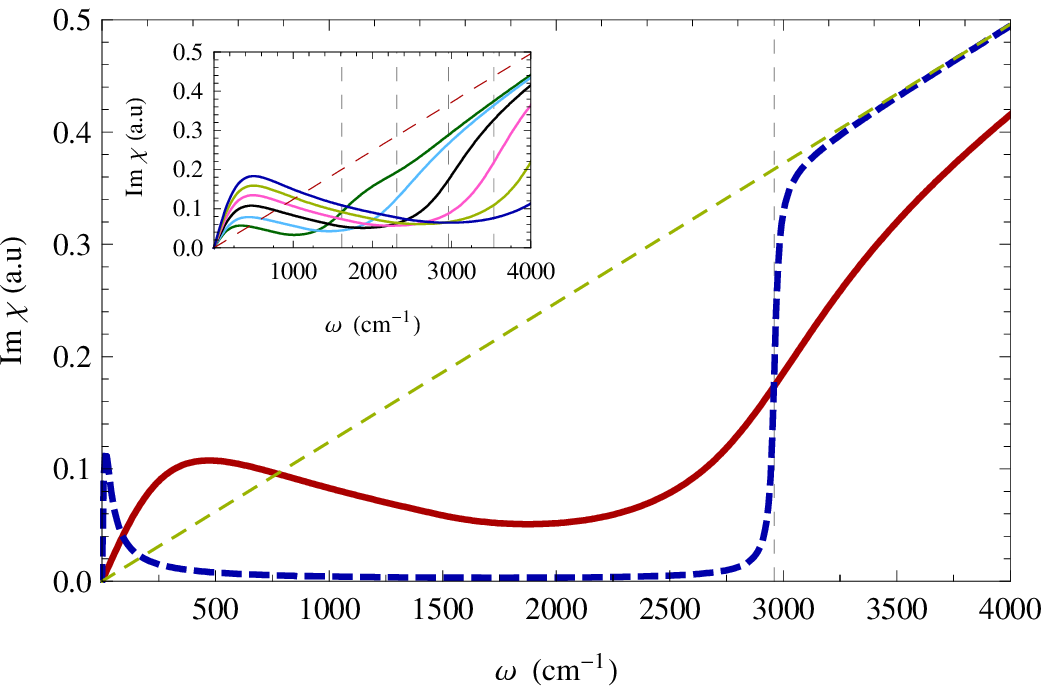}
\begin{center}
(b)
\end{center}
\end{center}
\end{minipage}\\
\caption{\label{fig2} (Color online) (a)  Comparison of the
optical conductivity computed with the self-energy (\ref{fig2})
for parameters $a=0.2$, $b=0.001$ eV, $\Lambda=0.15$ eV, $\mu=0.18$
eV (red solid line) and same with a constant scattering rate only i. e.
$a=0$ (blue dashed line). (b) Same comparison for the Raman signal. The
inset shows the evolution of the Raman signal obtained with the
self-energy (\ref{TwopartSE}) for decreasing values of the chemical
potential from top to bottom. The same color code as
in the optical conductivity is used. }
\end{figure*}
From a linearly dependent  imaginary part  of the electron
self-energy $\mathrm{Im}\;\Sigma(\omega)=a\vert\omega\vert$, and
computing the real part by a Kramers-Kroning transformation we get
the ``marginal Fermi liquid" type of the electron self-energy:
\begin{equation}
\label{SEMFL} \Sigma_{MFL}(\omega) =
\dfrac{2a}{\pi}\left[\omega\ln\left(\dfrac{\omega_{c}}{\omega}
\right) + i\dfrac{\pi}{2}|\omega| \right]+i\hspace{0.07cm} b,
\end{equation}
where we have also added a constant scattering rate. While in the
physics of cuprates  the electron self-energy shows a marginal
Fermi liquid behavior for a large range of doping
levels\cite{Vetal89}, in the case of graphene the marginal behavior
is restricted to a very small energy region around the zero doping
and chemical potential. We will take this fact into account by
considering the following  phenomenological electron self-energy:
\begin{equation}
\label{TwopartSE}
   \mathrm{Im}\;\Sigma(\omega) =
     \begin{cases}
       a|\omega |+b &\text{,} |\omega | < \Lambda \\
        a\Lambda + b & , |\omega | \geq \Lambda
     \end{cases},
\end{equation}
where $\Lambda$ is a low energy cutoff that will be set to
$\Lambda\sim 0.15$ eV. As mentioned before realistic values for
the parameters $a, b$ estimated from the experiment \cite{LLA09}
are $a=0.07$,  $ b=0.01$ eV although they may vary from sample
to sample.

A  Kramers-Kroning transformation of (\ref{TwopartSE}) gives for
the real part
\begin{equation}
\mathrm{Re}\;\Sigma(\omega) = \dfrac{a}{\pi}\left[
\Lambda\ln\left( \dfrac{\Lambda+\omega}{\Lambda - \omega}\right)
+ \omega\ln\left( \dfrac{\omega^{2}-\Lambda^{2}}{\omega^{2}}
\right)  \right].
\end{equation}
%

{\it Optical properties of the model.} From the Kubo formula we
can write the optical conductivity as a function of the electron
spectral function $A_{s}(k,\omega)$ in the so-called bubble
approximation that neglects the vertex corrections \cite{M93}:
\begin{eqnarray}
\label{bubble}
\mathrm{Re}\;\sigma(\omega)=g_{s}g_{v}\dfrac{e^{2}v^{2}_{F}}{4\omega}
\sum_{s,s'}\int \frac{k \;d k}{(2\pi)^{2}} . \nonumber \\ \int
d\epsilon \left[n_{F}(\epsilon) -
 n_{F}(\epsilon+\omega)\right]
A_{s}(k,\omega)A_{s'}(k,\epsilon+\omega),
\end{eqnarray}
where $g_{s,v}$ are the spin and valley degeneracies,
$n_{F}(\epsilon)=\left[\exp(\beta(\epsilon-\mu))+1\right]^{-1} $
is the Fermi distribution function, and $A_{s}({\bf k},\omega)$ is
the spectral function:
\begin{equation}
A_{s}(k,\omega) = \dfrac{-2\mathrm{Im}\;\Sigma(\omega,\mathbf{k})}
{(\omega -
sv_{F}|\mathbf{k}|-\mathrm{Re}\;\Sigma(\omega,\mathbf{k}))^2
+(\mathrm{Im}\;\Sigma(\omega,\mathbf{k}))^2}.
\end{equation}
The index $s=\pm1$ refers to the upper (lower) part of the
dispersion relation. The bubble approximation works well if the
electron self-energy is not strongly k-dependent \cite{M04} as
happens in our case.

Finally for systems with a very isotropic Fermi surface as it is
our case the electronic contribution to the Raman signal
$\mathrm{Im}\;\chi(\omega)$ can be related to the optical conductivity by the
simple expression \cite{SS90}:
\begin{equation}
\label{condRaman} \omega\;\mathrm{Re}\;\sigma(\omega) \sim\;
\mathrm{Im}\;\chi(\omega).
\end{equation}

\noindent
{\it Results and discussion.}
We have computed the optical conductivity from eqs.
(\ref{TwopartSE}) and (\ref{bubble}).

Fig. \ref{Basov} is the main result of this work. It shows a
comparison between the dynamical conductivity measured in ref.
\cite{Letal08} (a) and that obtained with the self-energy given by
eq. (\ref{TwopartSE}) (b) for the following values of the
parameters: $a = 0.2$, $ b = 0.001$ eV, $ \Lambda = 0.15$ eV. In
the experimental figure it can be seen a contribution coming from
de Drude peak at low energies, the interband transitions beyond
$2\mu$ and the strong background between the Drude peak and the
interband transitions. It can also be appreciated that the width
of the threshold of the interband transitions at $2\mu$ is very
wide and almost independent of the electronic density. In our
approach we obtain the most prominent characteristics of the
experimental results as can be appreciated in Fig. \ref{Basov}(b).

In order to clarify the role of the Coulomb interactions encoded
in parameter $a$ of (\ref{TwopartSE}) versus a simple constant
scattering rate $b$, we have shown in Fig. \ref{fig2} (a) a
comparison of the optical conductivity computed with the
self-energy (\ref{fig2}) for parameters $a=0.2$, $b=0.001$ eV,
$\Lambda=0.15$ eV, $\mu = 0.18$ eV (red solid line) and the same
with only a constant scattering rate (a=0) (blue dashed line). For
the constant scattering rate the intermediate frequency region
between Drude and interband transitions is almost zero and the
threshold is sharp. We have checked that the same kind of behavior
of the conductivity  is found using a Fermi liquid type of
self-energy which is expected to correspond to doped graphene
\cite{HGS07}. When the linear contribution to the scattering rate
is added the Drude peak becomes wider, there is a considerable
background at intermediate frequencies and the threshold of the
interband transitions acquires a finite width. All these
characteristics are expected from a linear scattering rate since
it induces a general shift of the spectral weight from lower to
higher frequencies. It is worth mentioning that although we have
added the cutoff $\Lambda$ in expression (\ref{TwopartSE}) to
restrict the marginal behavior to low frequencies, the shift of
the spectral weight is quite remarkable.

The problem of the anomalous residual
conductivity was studied in \cite{PSC08,Stetal08}
taking into account unitary scatterers, charge impurities and optical
and acoustic phonons. They arrive to the result
that to account for the observed value of the residual conductivity
another mechanism effective at high energies must be included.
They also obtain that optical phonons give a marginal contribution similar
to Coulomb interactions for frequencies away from
the Dirac point.
Some types of disorder also induce a
linear behavior on the scattering rate at low energies \cite{AS02}.

What we see from our analysis is that including a linear imaginary
part of the self-energy even for a very low energy range
stabilizes the background of the conductivity and widens the
threshold of the interband transitions in a way that is almost
independent of the electronic density. To get a better fit for the
experimental data we have chosen a big value of the slope of the
linear scattering rate ``$a$". We must mention that besides the
electronic interaction there can be other contributions to the
linear scattering rate coming from disorder \cite{AS02} or optical
phonons \cite{PSC08,Stetal08}.


Fig. \ref{fig2}(b) shows the same comparison for the electronic
contribution to the Raman signal in graphene yet to be measured.
Dashed blue curve refers to the case with a constant scattering
rate. The electronic Raman scattering has been calculated very
recently in Ref. \cite{KF09}. In agreement with our work they also
find a linear spectral density beyond $2\mu$ coming from interband
transitions. We obtain a Drude contribution at low energy coming
from the constant scattering rate $b=0.001$ eV (dotted blue line).
As happens in the optical conductivity  there is no spectral
weight in the region between intra and interband transition when
only a constant scattering rate is considered. Adding the marginal
like behavior however (red solid line) there is a noticeable
background at intermediate frequencies and there are no sharp
separations between the Drude contribution, the background and the
interband transitions. The inset of Fig. \ref{fig2}(b) shows the
variation of the Raman spectrum with the chemical potential. The
values plotted are the same as in the optical conductivity. The
three main characteristics i.e. : Drude contribution, interband
transitions and background are clearly seen. For lower chemical
potentials it can be seen that the Drude peak shifts to higher
frequencies when increasing $\mu$ since the linear contribution
dominates the physics. This behavior will also be present in the
Drude contribution of the optical conductivity but such low
frequencies are not accessible in optical conductivity
experiments.

To summarize, we have proposed a phenomenological model that
includes the linear behavior of the scattering rate at low
frequencies and shown that it can explain qualitatively recent
experiments in optical conductivity of graphene \cite{Letal08}.
The linear scattering rate might have its origin on electronic
interactions as discussed in this paper but   optical phonons and
disorder can also contribute to strengthen the marginal Fermi
liquid type of behavior. In particular, we have obtained the
strong background observed between the Drude peak and interband
transitions and the wide width of the threshold of the interband
transitions, two features that are difficult to explain using a
Fermi liquid picture. We have predicted that an anomalous
background will be also seen in other optical probes such as
electronic Raman scattering.
\\

{\it Acknowledgments} We acknowledge very useful conversations
with Fernando de Juan and Javier Sabio. We also thank  Dimitri
Basov for kindly allow us to reproduce Fig. 2 (b) of ref.
\cite{Letal08}. This research was supported by the Spanish MECD
grant FIS2005-05478-C02-01 and FIS2008-00124.

\bibliography{Marginal}

\end{document}